\documentclass[epj]{svjour}
\pdfoutput=1

\usepackage{amssymb,amsmath}
\usepackage{graphicx}
\usepackage{color}
\usepackage{soul}
\graphicspath{{figures/}}

\usepackage{verbatim}

\begin{document}

\newcommand{\bmath}{\begin{displaymath}}
\newcommand{\emath}{\end{displaymath}}

\newcommand{\be}{\begin{equation}}
\newcommand{\ee}{\end{equation}}
\newcommand{\bea}{\begin{eqnarray}}
\newcommand{\eea}{\end{eqnarray}}
\newcommand{\non}{\nonumber\\}
\newcommand{\bmultl}{\begin{multline}}
\newcommand{\emultl}{\end{multline}}

\newcommand{\bsubeq}{\begin{subequations}}
\newcommand{\esubeq}{\end{subequations}}
\newcommand{\bitemize}{\begin{itemize}}
\newcommand{\eitemize}{\end{itemize}}
\newcommand{\ket}[1]{\left|{#1}\right\rangle}
\newcommand{\bra}[1]{\left\langle{#1}\right|}
\newcommand{\abs}[1]{\left|{#1}\right|}
\newcommand{\re}{\mathrm{Re}}
\newcommand{\im}{\mathrm{Im}}
\newcommand{\bmx}{\begin{bmatrix}}
\newcommand{\emx}{\end{bmatrix}}
\newcommand{\bsmx}{\begin{smallmatrix}}
\newcommand{\esmx}{\end{smallmatrix}}

\def\ud{\mathrm{d}}
\def\dt{\frac{\partial}{\partial t}}
\def\R{\vec{\hat{R}}}
\renewcommand{\Re}{\mathrm{Re}}
\renewcommand{\vec}[1]{\underline{#1}}
\newcommand{\mathbold}[1]{\mathbf{#1}}
\newcommand{\rev}[1]{\vec{r}^{(#1)}}
\newcommand{\lev}[1]{\vec{l}^{(#1)}}

\def\wR{\omega_R}
\def\da{\delta{}a}
\def\db{\delta{}b}
\def\dc{\delta{}c}

\newcommand{\bquote}{\quotedblbase{}}
\newcommand{\equote}{\textquotedblright{ }}

\graphicspath{{./figures/}}

\title{Cavity optomechanics with a trapped, interacting Bose-Einstein condensate}
\author{D.~Nagy\thanks{\email{nagy.david@wigner.mta.hu}} \and
  G.~Szirmai \and P.~Domokos }                     
\institute{Institute for Solid State Physics and Optics, Wigner
  Research Centre for Physics, Hungarian Academy of Sciences,
  P. O. Box 49, H-1525 Budapest, Hungary}
\date{Received: date / Revised version: date}
%

\abstract{The dispersive interaction of a Bose-Einstein condensate
  with a single mode of a high-finesse optical cavity realizes the
  radiation pressure coupling Hamiltonian. In this system the role of
  the mechanical oscillator is played by a single condensate
  excitation mode that is selected by the cavity mode function. We
  study the effect of atomic s-wave collisions and show that it merely
  renormalizes parameters of the usual optomechanical
  interaction. Moreover, we show that even in the case of strong
  harmonic confinement---which invalidates the use of Bloch states---a
  single excitation mode of the Bose-Einstein condensate couples
  significantly to the light field, that is the simplified picture of
  a single ``mechanical'' oscillator mode remains valid.
}

\maketitle

\section{Introduction}

Cavity optomechanics has seen a rapid expansion over recent years
attracting great theoretical and experimental interest. The basic
paradigm of coupling a vibrational mode of a mesoscopic object to an
electromagnetic field mode via radiation pressure force has been
realized in various systems (for a comprehensive review consult
\cite{Meystre2012Short}). With nanomechanical oscillators, it still
remains a challenge in the optical domain to reach simultaneously the
quantum mechanical ground state {\it and} the quantum coherent regime,
where the coupling exceeds both the optical and mechanical decoherence
rates \cite{Verhagen2012Quantumcoherent}. In contrast, these goals are
relatively easily achieved with ultracold atoms
\cite{StamperKurn2012Cavity,Purdy2010Tunable,Brooks2012Nonclassical}. Most
remarkably, radiation pressure coupling can be simulated with a
Bose-Einstein condensate (BEC) dispersively coupled to the field of a
high-finesse optical cavity, that can be either a ring cavity
\cite{Slama2007Superradiant,Slama2007Cavityenhanced,Steinke2011Role,Zhang2012Role}
or a linear microcavity
\cite{Brennecke2007Cavity,Brennecke2008Cavity}. In the latter case,
the cavity mode function selects a single excitation mode of the BEC,
which plays the role of the mechanical oscillator.  For a pure
condensate the occupation of the selected excitation is already zero,
thus one can readily study the quantum coherent regime of
optomechanics. As compared to the single atom cavity QED, the coupling
of the collective excitation mode to the field is enhanced by the
square root of the atom number. In this strong coupling regime optical
nonlinearity emerges on a single photon level
\cite{Gupta2007Cavity,Ritter2009Dynamical}, which may give rise to
photon blockade effect \cite{Rabl2011Photon} and non-Gaussian steady
states \cite{Nunnenkamp2011SinglePhoton}. This regime also opens up
the way to possible future applications of BEC--cavity systems as
optical transistors \cite{Chen2011Alloptical} and switches
\cite{Yang2011Controllable}.

Nonetheless, using a BEC excitation mode as a ``mechanical''
oscillator sets important limitations. The frequency of this
matter-wave oscillator, defined by the recoil frequency $\omega_R$, is
by orders of magnitudes lower than the coupling strength and usually
than the cavity photon loss rate $\kappa$. As a result, the quantum
noise associated with the photon leakage gives rise to significant
incoherent excitations of the BEC. The rate of diffusion out of the
ground state is comparable to the interaction strength, which leads to
the dephasing of coherent oscillations \cite{Nagy2009Nonlinear}. On a
longer time scale, the condensate depletes into the ``mechanical''
oscillator mode which approaches a steady state with average
occupation number being in the order of $\kappa/\omega_R$
\cite{Szirmai2009Excess,Szirmai2010Quantum}.

In this article, we shall study two further effects which are inherent
to the BEC experimental approach to optomechanics and which might lead
to significant departure from the realization of the radiation
pressure coupling models. The first one is the effect of internal
interactions on the collective excitation mode. Indeed, the effect of
s-wave atom-atom collisions cannot be generally neglected, as it
plays, e.g., a crucial role in the process of Bose-Einstein
condensation, too. In a recent paper the collisional effects have been
considered quite generally for the BEC-cavity system
\cite{Dalafi2013Nonlinear}. Here we find for the special case of the
simulated radiation pressure model that the internal interaction
renormalizes the model parameters, however, the generic form remains
good approximation. The second effect is due to the external
trapping. Since the condensate is inhomogeneous and has a finite
extent of a few wavelengths, the selection of a single ``mechanical''
mode is not perfectly warranted. We find that the single-mode
approximation to optomechanics can be applied already for a trap size
as small as 10 times the optical wavelength.

The paper is organized as follows: In Sec.~\ref{sec:beccav} we
introduce the dispersively coupled cavity and atom
fields. Sec.~\ref{sec:untrapped} describes the cavity optomechanics with
a homogeneous BEC. We develop a band model for the periodic system,
and show that the cavity mode  singles one excitation mode out of the
continuum. We describe the effects of collisions
on the dispersive bistability of the system both by a full mean-field
model and by an effective radiation pressure coupling
Hamiltonian. Sec.~\ref{sec:trapped} investigates the limitations of
optomechanics for a trapped BEC. Finally, we conclude in
Sec.~\ref{sec:conclusion}.

\section{The BEC-cavity system}
\label{sec:beccav}

We consider a zero-temperature Bose-Einstein condensate trapped inside
a high-finesse optical resonator. The atoms interact with a single
cavity mode having frequency $\omega_C$ and mode function $\cos\,kx$,
where $k=2\pi/\lambda$ is the wave number. The system is driven
through one of the cavity mirrors by a coherent laser field of
frequency $\omega$ which is far detuned from the atomic transition
$\omega_A$, {\it i.e.}, the atomic detuning $\Delta_A=\omega -
\omega_A$ far exceeds the atomic linewidth $\gamma$, hence spontaneous
emission is suppressed. Meanwhile, the laser field is nearly resonant
to the cavity mode, $|\Delta_C| = \omega - \omega_C \approx \kappa$,
with $2\kappa$ being the cavity mode linewidth. We assume a strong
dipole coupling between the atoms and the mode characterized by the
single-photon Rabi frequency $\Omega$ which is in the order of the
dissipation rates $\kappa$, $\gamma$. In this limit the strength of
the dispersive atom-field interaction is defined by the one-atom light
shift $U_0 = \frac{\Omega^2}{\Delta_A}$
\cite{Domokos2003Mechanical}. In the frame rotating with the laser
frequency $\omega$, the many-particle Hamiltonian reads \cite{Szirmai2010Quantum}
\begin{multline}
\label{eq:H_total}
H/\hbar = -\Delta_C\,a^\dagger{}a + \eta(a + a^\dagger) \\ \qquad\qquad
 + \int \Psi^\dagger(x)\bigg[-\frac{\hbar}{2\,m}\frac{d^2}{dx^2}+ V_{\rm ext}(x)
   \\\qquad\qquad  + \frac{g}{2}\Psi^\dagger(x)\Psi(x)   +
   U_0\,a^\dagger{}a\cos^2(kx) \bigg]\Psi(x)dx\,,
\end{multline}
where $\Psi(x)$ is the atom field operator and $a$ is the cavity mode
operator. We consider the dynamics in one dimension $x$ along the
cavity axis.  The first two
terms describe the radiation field and the pumping of the cavity
mode. The energy scale of the atomic motion is given by the recoil
frequency $\omega_R = \hbar{}k^2/(2m)$. The external
trapping potential  $V_{\rm ext}(x)$ is assumed to vary slowly on the 
wavelength scale.  We include s-wave atom-atom collisions with the 1D
interaction parameter $g$, which is proportional to the s-wave
scattering length. The last term accounts for the dispersive
light-matter interaction between the atoms and the mode.

In cavity optomechanics one has to consider a dissipative dynamics
for the photon field due to the photon leakage through the cavity
mirrors. We take this into account in the Heisenberg-Langevin
equations of motion of the field operator \cite{Szirmai2009Excess,Szirmai2010Quantum}
\begin{equation}
\label{eq:cavity}
  \frac{d}{d t} \hat a = -\frac{i}{\hbar} \left[\hat a, H \right] -
  \kappa \hat a + \hat \xi\,,  
\end{equation}
where $2\kappa$ is the photon loss rate and the operator $\xi$
describes the measurement back-action noise with the only non-zero
correlation function $\langle \hat \xi(t) \hat \xi^\dagger (t')
\rangle = \kappa \delta(t-t')$. 

\section{Optomechanical coupling in a band model}
\label{sec:untrapped}

The system can be considerably simplified when the condensate is
homogeneous. In this case, the cavity field couples exclusively to a
\emph{single} excitation mode of the BEC, thus an effective
radiation pressure Hamiltonian can be constructed for these two modes. The effect of
atom-atom collisions is that they renormalize the optomechanical
coupling strength and the frequency of the ``mechanical'' mode.

In the absence of an external potential, $V_{\rm ext}(x)
\equiv 0$, the problem is periodic along the cavity axis. The
interaction term in Hamiltonian (\ref{eq:H_total}) has the periodicity
of $\lambda/2$, where $\lambda$ is the optical wavelength. It follows
that one can introduce a band model  \cite{Szirmai2010Quantum,Venkatesh2011Band-structure} by expanding the atomic field
operators in terms of Bloch functions,
\begin{multline}
\label{eq:mode_exp}
\Psi(x) = \frac{1}{\sqrt{L}}
\sum_{q}e^{iqx} \\
\left\{b_q + \sqrt{2}\sum_n\left[c_{n,q}\cos\,2nkx + s_{n,q}\sin\,2nkx \right]\right\}\,,
\end{multline}
where $n$ is the band index and $q\in [-k, k]$ denotes the
quasimomentum in the first Brillouin zone, and $L$ is the linear size of the system. $b_q$, $c_{n,q}$ and
$s_{n,q}$ are annihilation operators of the corresponding states.
We use the standing wave basis $e^{iqx}\cos\,2nkx$, and $e^{iqx}\sin\,2nkx$ instead of plane waves because the atom-field interaction directly populates  the cosine wave functions with $q=0$. 

The energy of the bands depend quadratically on the band index,
$E_{n,q=0} = (2n)^2\omega_R$, hence there is a hierarchy in their
population as shown in Ref.~\cite{Konya2011Multimode}. We truncate the
expansion Eq.~(\ref{eq:mode_exp}) at $n=1$, meaning that we make a
three-band approximation and restrict the dynamics in the three lowest
lying bands $b_q$, $c_{1,q} \rightarrow c_{q}$ and $s_{1,q}
\rightarrow s_{q}$. Accordingly,
\be
\label{eq:atomnum}
\int_0^L\Psi^\dagger(x)\Psi(x)dx = \sum_q\left[b_q^\dagger{}b_q +
  c_q^\dagger{}c_q + s_q^\dagger{}s_q\right] = N \,.  
\ee 
This is valid when the optical potential depth $U_0\langle
a^\dagger{}a\rangle$ is below the recoil frequency.

By inserting Eq.~(\ref{eq:mode_exp}) into Hamiltonian~(\ref{eq:H_total}) (but keeping only $n=1$ from the sum),
we obtain the following second-quantized form,
\begin{multline}
\label{eq:H_quantized}
H = -\delta_C\,a^\dagger{}a +
  \eta(a + a^\dagger) +\frac{U_0}{2\sqrt{2}}\, a^\dagger{}a \sum_q
  \left(b_q^\dagger{}c_q + c_q^\dagger{}b_q \right) \\ + 
  H_{\rm kin} + H_{\rm coll}\,,
\end{multline}
where we introduced the shifted cavity detuning $\delta_C = \Delta_C -
\frac{NU_0}{2}$. The atom-cavity interaction (third term) couples the
$b_q$ modes to the $c_q$ modes preserving the quasi momentum $q$. The kinetic energy reads
\begin{multline}
\label{eq:H_kin}
 H_{\rm kin} = \omega_R \sum_q \Big[ q^2 b_q^\dagger{}b_q + (q^2 +
    4)(c_q^\dagger{}c_q + s_q^\dagger{}s_q) \\ + 4iq(c_q^\dagger{}s_q -
    s_q^\dagger{}c_q)\Big] \,.
  \end{multline}
Note that the sine modes $s_q$, not present in the atom-cavity coupling (third term in Eq.~(\ref{eq:H_quantized})), 
mix with the cosine modes $b_q$ here. They are thus needed to recover the correct excitation spectrum.
Finally, the collisions are represented by
\begin{multline}
\label{eq:H_coll}
  H_{\rm coll} = 
  \frac{g}{2}\frac{\lambda}{L}\sum_{q_1,q_2,q_3,q_4} \delta_{q_1+q_2 -
    q_3 - q_4} \Big[ 
    \\ 
    b_{q_1}^\dagger{}b_{q_2}^\dagger{}b_{q_3}{}b_{q_4} +
    \frac{3}{2}c_{q_1}^\dagger{}c_{q_2}^\dagger{}c_{q_3}{}c_{q_4} +
    \frac{3}{2}s_{q_1}^\dagger{}s_{q_2}^\dagger{}s_{q_3}{}s_{q_4}
    \\ + 
    b_{q_1}^\dagger{}b_{q_2}^\dagger{}c_{q_3}{}c_{q_4} +
    b_{q_1}^\dagger{}b_{q_2}^\dagger{}s_{q_3}{}s_{q_4} + \frac{1}{2}
    c_{q_1}^\dagger{}c_{q_2}^\dagger{}s_{q_3}{}s_{q_4} + \mbox{h.a.}
    \\ + 
    4b_{q_1}^\dagger{}c_{q_2}^\dagger{}b_{q_3}{}c_{q_4} +
    4b_{q_1}^\dagger{}s_{q_2}^\dagger{}b_{q_3}{}s_{q_4} +
    2c_{q_1}^\dagger{}s_{q_2}^\dagger{}c_{q_3}{}s_{q_4} + \mbox{h.a.} \Big]\,.
\end{multline} 
The s-wave atom-atom scattering mixes operators
with different $n$ and $q$ preserving the total momentum. In the
above Hamiltonian, we neglected Umklapp scattering processes, where
the band momentum is transformed into quasi momentum, and considered
only normal scattering where the quasi momentum is preserved,
i.e. $q_1+q_2-q_3-q_4 = 0$.

\subsection{Mean-field description}
\label{subsec:mf_model}

In order to describe the coupled dynamics of the BEC and the photon field we employ the same technique \cite{Szirmai2010Quantum} that leads to the Gross-Pitaevskii equation for a Bose gas. Hence, the
operators are split into their mean values and fluctuations,
\begin{subequations}
  \label{eq:mf_expansion}
\begin{align}
a(t) &= \alpha + \delta{}a \,,\\
b_q(t) &= e^{-i\mu{}t}\left[\sqrt{N}\beta_q + \delta{}b_q\right] \,,\\
c_q(t) &= e^{-i\mu{}t}\left[\sqrt{N}\gamma_q + \delta{}c_q\right] \,,\\
s_q(t) &= e^{-i\mu{}t}\left[\sqrt{N}\sigma_q + \delta{}s_q\right]\,.  
\end{align}
\end{subequations}
For the atoms the mean field can be interpreted as the wave
function of the condensate with components $\beta_q$, $\gamma_q$ and
$\sigma_q$ in the corresponding states, and chemical potential $\mu$.
Note that the wave function can be chosen real.  Momentum is conserved
simultaneously for the mean field and for the fluctuations.  If
initially a homogeneous condensate is prepared in the zero momentum
state with only $\beta_0 \neq 0$, then due to momentum conservation
the condensate wave function stays in the $q=0$ subspace even after adiabatically turning on the cavity field.  Furthermore
the parity is conserved, therefore the atom-cavity interaction
populates only $\gamma_0$, and leaves $\sigma_0 = 0$.

One obtains the following three coupled equations in the steady state for the mean fields by taking the average of the Heisenberg-Langevin equation of motion [Eq. \eqref{eq:cavity} and similarly for the atomic fields from the Hamiltonian \eqref{eq:H_quantized}]:
\begin{subequations}
\label{eq:tm_mf}
\be
\alpha = \frac{\eta}{i\kappa + (\delta_C - 2Nu\beta_0\gamma_0)}  \,,
\ee
\be
\mu\beta_0 = u |\alpha|^2 \gamma_0 + {\cal G}\left(\beta_0^3 +
3\beta_0\gamma_0^2\right) \,,
\ee
\be
\mu \gamma_0 = 4\omega_R\gamma_0 + u |\alpha|^2 \beta_0 + {\cal G}\left(\frac32
\gamma_0^3 + 3 \beta_0^2\gamma_0 \right) \,,
\ee
\end{subequations}
together with the normalization $\beta_0^2 + \gamma_0^2 = 1$. For
brevity, we introduced the notations $u = \frac{U_0}{2\sqrt{2}}$ and the
collisional parameter ${\cal G} = gN\lambda/L$. 

The equation of motion of the fluctuations are linearized around the
mean field solution. As a result of momentum conservation the
equations decouple for distinct quasimomenta. S-wave scattering mixes the
$\pm{}q$ quasimomentum states, hence, for each $q\neq 0$, one has to
solve  six coupled equations.  This is precisely the Bogoliubov problem for a BEC in a periodic potential (restricted to the lowest three bands). By arranging the variables into 
 $\vec{R}_q = [\delta{}b_q,
  \delta{}b_{-q}^\dagger, \delta{}c_q, \delta{}c_{-q}^\dagger,
  \delta{}s_q, \delta{}s_{-q}^\dagger]^T$, the following compact
notation is obtained for $q \neq 0$\,:
\be
\label{eq:fluct_q}
i\frac{d}{dt} \vec{R}_q = {\bf L}_q \vec{R}_q \,,
\ee
where ${\bf L}_q$ is a six by six matrix whose elements
depend on the mean-field solution and the quasi momentum $q$ (for the
complete form of ${\bf L}_q$ see the Appendix). 
The $q=0$ fluctuations have to be treated differently. Within
the mean-field model, the cavity field fluctuations ($\delta a$)
couple only to the even parity states with $q=0$ ($\delta b_0$ and 
$\delta c_0$). These operators can be arranged accordingly into the 
vector $\vec{R} = [\delta{}a, \delta{}a^\dagger, \delta{}b_0, \delta{}b_0^\dagger,
  \delta{}c_0, \delta{}c_0^\dagger]^T$, and the coupled BEC-cavity
excitations are described by 
\be
\label{eq:fluct_0}
i\frac{d}{dt} \vec{R} = {\bf M} \vec{R} + \vec{\xi}\,,
\ee
with ${\bf M}$ being a six by six complex matrix that depends on
the mean field solution (for its complete form see the Appendix). The
cavity field fluctuations are driven by the dissipation noise that is
arranged into the vector $\vec{\xi} = [\xi, \xi^\dagger, 0, 0, 0,
  0]^T$.
\begin{figure}[t]
\centering
\includegraphics[width=0.8\columnwidth]{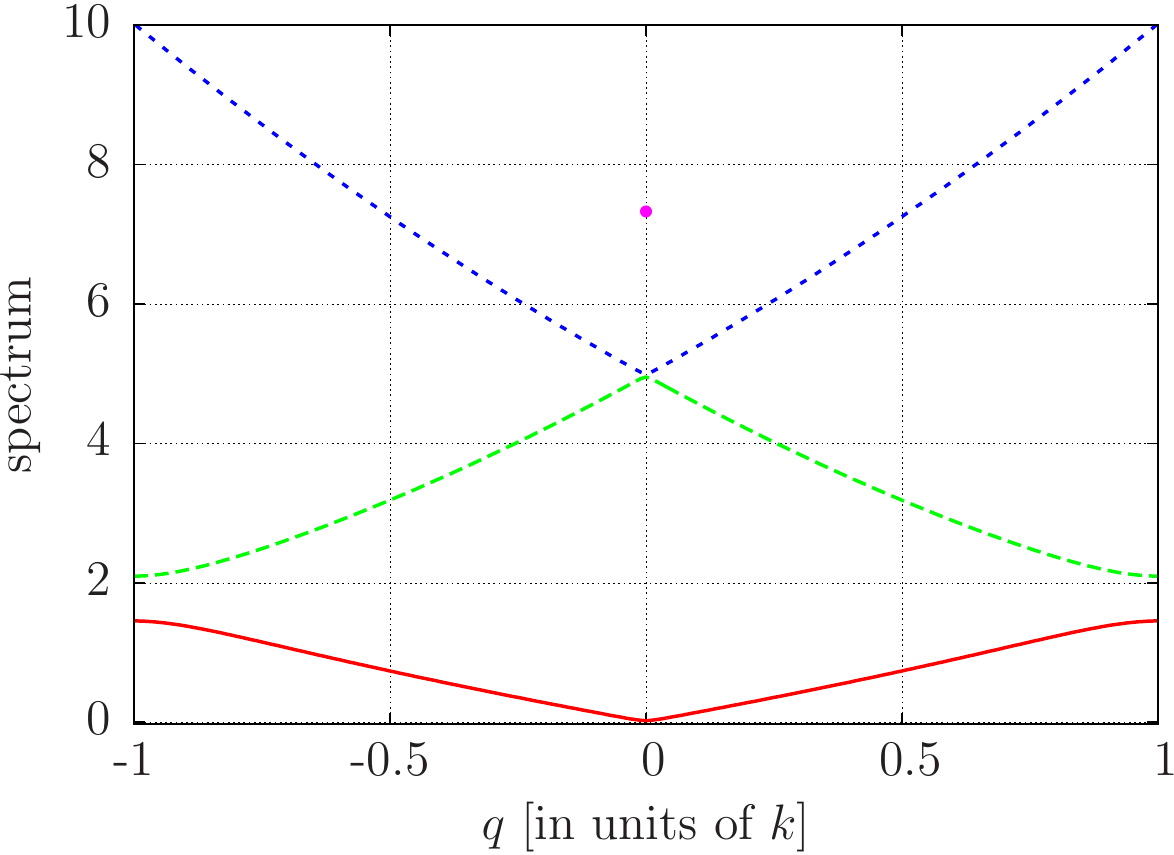}
\caption{Band structure of the BEC excitation spectrum. The energy
  eigenstates for $q\neq 0$ depends on the cavity field only through
  the mean photon number. The optical potential opens gaps between
  adjacent bands. A single $q=0$ state (filled circle) is pulled out
  from the third band (dashed blue) by the interaction with cavity
  field fluctuations. The parameters are $N=6\cdot 10^4$, $U_0 =
  0.96$, $\eta=549.5$, $\kappa=363.9$, $\delta_C = -5120$ and ${\cal
    G} = 1$ $\omega_R$.}
\label{fig:bands}
\end{figure}

The excitation spectrum of the BEC interacting with the cavity field
is depicted in Fig.~\ref{fig:bands}. The periodic optical potential
leads to a band structure, as it opens gaps at the center and at the
edges of the Brillouin zone. In the first band, the spectrum starts
linearly in $q$, that is typical to the Bogoliubov spectrum of a
homogeneous BEC. Interaction with the photon field fluctuations
picks out a single mode with $q=0$ from the upper band, which constitutes
the optomechanical mode. In the following we will concentrate on this
relevant mode, and derive an effective radiation pressure Hamiltonian
that includes s-wave collisions.

\subsection{Radiation pressure Hamiltonian}
\label{subsec:optomcoupling}

Radiation pressure coupling between the cavity mode $a$ and the single BEC
excitation mode $c_0$ can be obtained from
Hamiltonian~(\ref{eq:H_quantized}) by restricting the dynamics in the
subspace of the modes $b_0$ and $c_0$.

The Hamiltonian in the restricted subspace (up to second order in
$c_0$ and $c_0^\dagger$) reads
\begin{multline}
 \label{eq:hamtwomodes}
H = -\delta_C a^\dagger{}a + 4\omega_R c_0^\dagger c_0 + u a^\dagger{}a
\left(b_0^\dagger{}c_0 + c_0^\dagger{}b_0 \right) \\
+ \frac{g}{2}\frac{\lambda}{L} \left(
b_0^\dagger{}b_0^\dagger{}c_0{}c_0 +
c_0^\dagger{}c_0^\dagger{}b_0{}b_0 + 4
b_0^\dagger{}c_0^\dagger{}b_0{}c_0 +
b_0^\dagger{}b_0^\dagger{}b_0{}b_0 \right)\,.
\end{multline} 
First let us exploit the relation $b_0^\dagger b_0 = N - c_0^\dagger c_0$ (Eq.~(\ref{eq:atomnum})) in the last term of Eq.~\eqref{eq:hamtwomodes}, and then apply the next level of approximation: $b_0 \equiv \sqrt{N}$. By this we assume that the condensate is mainly homogeneous, and the relevant dynamics takes place in the $c_0$
mode. This approximation is valid in the limit $u|\alpha|^2 \ll 4\omega_R$.
Keeping only the quadratic terms from the s-wave collisions, one ends
up with the optomechanical Hamiltonian up to second order 
in the operators,
\be
\label{eq:H_om}
H = -\delta_C a^\dagger{}a + \frac{\omega_M}{2}(X^2 + Y^2) + G
a^\dagger{}a X + {\cal G} X^2 \,,
\ee
where we introduced the quadratures $X = (c_0^\dagger +
c_0)/\sqrt{2}$, $Y = (c_0^\dagger + c_0)/\sqrt{2}$. The ``mechanical''
frequency is $\omega_M = 4\omega_R$ and the coupling constant is $G =
\sqrt{2N}u$. The s-wave interaction acts on the $X$ quadrature in the
last term, that can be eliminated by the Bogoliubov transformation
$\widetilde X = \chi X$ and $\widetilde Y = Y / \chi$. With the new operators,
we obtain the usual optomechanical Hamiltonian
\be
\label{eq:H_om_eff}
H = -\delta_C a^\dagger{}a + \frac{\widetilde{\omega}_M}{2}(\widetilde{X}^2 +
\widetilde{Y}^2) + \widetilde{G} a^\dagger{}a \widetilde{X}\,,
\ee
where the transformation parameter $\chi$ depends on the strength of
the s-wave interaction as $\chi = \sqrt[4]{(\omega_M + 2{\cal G})/\omega_M}$,
and it rescales both the mechanical frequency 
\begin{subequations}
\be
\label{eq:frequency_renormalized}
\widetilde\omega_M = \sqrt{\omega_M(\omega_M + 2{\cal G})}\,,
\ee
and the coupling constant 
\be \widetilde{G} = G/\chi\,. \ee
\end{subequations}
This is one of the main results of the present paper: the presence of atom-atom collisions besides shifting the oscillator frequency also renormalizes the optomechanical coupling.
In the regime of weak collisions, where ${\cal G}\ll \omega$, one
obtains from the Taylor series a linear frequency shift
$\widetilde{\omega}=\omega + {\cal G}$ and a coupling constant
$\widetilde{G}=\left(1 - {\cal G}/(2\omega)\right)G$.  The s-wave
collision between the atoms increases thus the mechanical frequency and
decreases the coupling strength.

\subsection{Effects of collisions on the dispersive bistability}
\label{subsec:bistability}

The nonlinear coupling between the BEC and the cavity mode results in a
bistable behavior of the system, that is one has two stable mean-field solutions corresponding to different photon numbers and oscillator displacements. This is a key feature of radiation
pressure coupling. In the following we compare the mean-field model,
Eqs.~(\ref{eq:tm_mf}a-c) to the mean-field solution of the effective
optomechanical model. This latter one is obtained from the
steady-state solution of Eq.~(\ref{eq:H_om_eff}) and
Eq.~(\ref{eq:cavity}) for the mean fields. By expressing $\langle X
\rangle$, one obtains the following cubic equation for the mean cavity
photon number $I=|\alpha|^2$,
\begin{figure}[t]
\centering
\includegraphics[width=0.8\columnwidth]{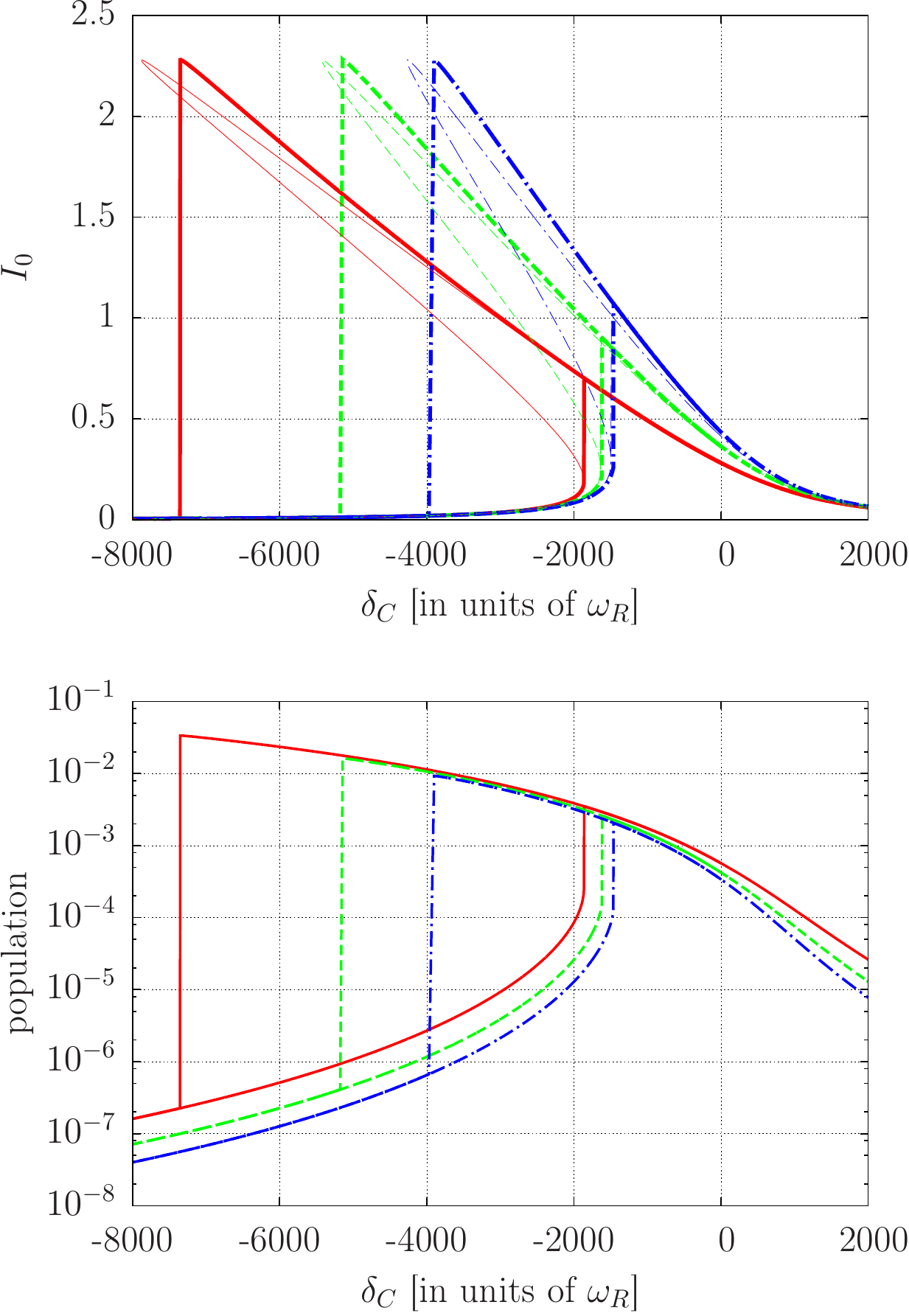}
\caption{Effects of collisions on bistability. Intracavity photon
  number $I_0$ (top panel) and the population of the optomechanical
  mode $|\gamma_0|^2$ (bottom panel) vs the detuning parameter
  $\delta_C$.  From the collisionless case (solid red line), the
  s-wave collision parameter is increased to ${\cal G} = 1$ (dashed
  green line) and ${\cal G} = 2$ (dashed dotted blue line). On the top
  panel the thick lines represent the solutions of
  Eqs.~(\ref{eq:tm_mf}a-c) while the thin lines show all of the three
  solutions of Eq.~(\ref{eq:cubic}).  The parameters reflects the
  experimental situation of Ref.~\cite{Brennecke2008Cavity}, $N=6\cdot
  10^4$, $U_0 = 0.96$, $\eta=549.5$, $\kappa=363.9\omega_R$.}
\label{fig:bistab}
\end{figure}
\be
\label{eq:cubic}
\frac{\widetilde{G}^4}{\widetilde{\omega}^2} I^3 + 2 \delta_C
\frac{\widetilde{G}^2}{\widetilde{\omega}} I^2 + (\delta_C^2 + \kappa^2)
I - \eta^2 = 0 \,, 
\ee 
Generally, this equation has one real and two complex solutions.
However above a bistability threshold  $\eta^2 \geq
\frac{8}{3\sqrt{3}}\frac{\widetilde{\omega}\kappa^3}{\widetilde{G}^2}$,
there is a finite interval in $\delta_C$ for which it has three real
solutions. One of them is unstable and the other two are stable, which
amounts to optical bistability. In the top panel of
Fig.~\ref{fig:bistab} we plot the solutions of Eq.~(\ref{eq:cubic})
(with thin lines) and compare them to the solutions of the full
mean-field model Eqs.~(\ref{eq:tm_mf}a-c) (thick lines) for three
different collision parameters ${\cal G} = 0$ (solid red), ${\cal G} =
1$ (dashed green) and ${\cal G} = 2$ (dashed dotted blue).  As ${\cal
  G}$ increases, the width of the bistable region decreases, while its
boundaries move upward in $\delta_C$. It is clearly seen that for photon numbers higher than $1$, the
optomechanical model gets less exact quantitatively. The underlying reason is that this model discards the depletion of the mode $b_0$. The bottom panel of
Fig.~\ref{fig:bistab} shows the mean-field occupation of the
optomechanical mode $|\gamma_0|^2$. One can observe that the s-wave
interaction reduces the population $|\gamma_0|^2$, thus smearing out the density modulation above the
homogeneous atom cloud. The decreasing $|\gamma_0|^2$ pulls the cavity closer to resonance
(cf. Fig.\ref{fig:bistab}).  Correspondingly, 
the photon number increases with ${\cal G}$ at a given $\delta_C$ on the upper
branch.

\begin{figure}[t]
\centering
\includegraphics[width=0.8\columnwidth]{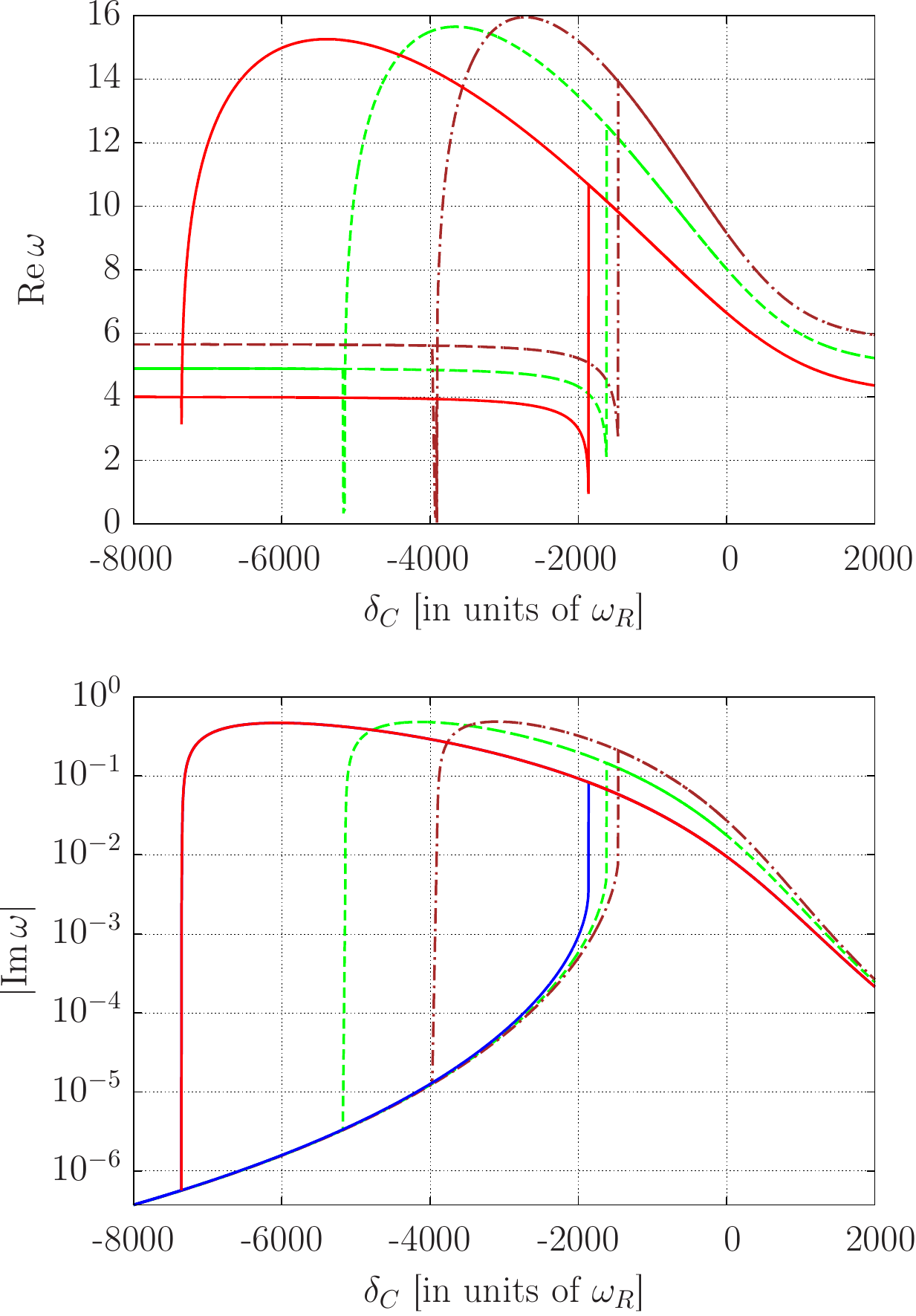}
\caption{Real part (top) and the absolute value of the imaginary part
  (bottom) of the optomechanical mode frequency as a function of the
  cavity detuning for ${\cal G}=0$ (solid red and blue lines), $1$
  (dashed green line) and $2$ (dashed dotted brown line). Other
  parameters are the same as in Fig.~\ref{fig:bistab}.}
\label{fig:optomode}
\end{figure}

In this section we focus on the behavior of the optomechanical mode (represented by the filled circle in Figure \ref{fig:bands}) across the
bistable regime. Its frequency is significantly modulated by the
interaction with the cavity field as shown in Fig.~\ref{fig:optomode}
(top panel). Far from the resonance, the frequency tends to $4\omega_R$ for ${\cal G}=0$ (solid red line), and to higher values given by Eq.~(\ref{eq:frequency_renormalized}) for  ${\cal G}\neq0$. In the bistability regime obtained at the vicinity of the cavity resonance, 
the optomechanical mode frequency is decreased  by the
BEC-cavity interaction on the lower branch, while it is increased on the upper branch. At the boundaries of
the bistable regime, the frequency that corresponds to the vanishing
branch sharply drops down to zero. The dissipative nature of the
cavity mode gives rise to cooling ($\mathrm{Im}\,\omega<0$, lower branch) or heating ($\mathrm{Im}\,\omega>0$, upper
branch) of the optomechanical mode \cite{Szirmai2010Quantum}. On the bottom panel of
Fig.~\ref{fig:optomode} we plot the absolute value of the imaginary part of the complex frequency
in logarithmic scale. (For ${\cal G}=0$ (solid line) we indicate cavity
cooling with blue, and cavity heating with red colors.)

\section{Trapped condensate}
\label{sec:trapped}

So far we have seen that a single excitation mode is sufficient to
seize most of the physics of a BEC-cavity optomechanical system. Now
we turn to the central issue of this article, that is, under what
conditions the single-mode effective optomechanical model can be valid
for a \emph{trapped} atom gas. In the experiments
\cite{Brennecke2007Cavity,Brennecke2008Cavity}, the condensate is kept
inside the cavity mode volume with an extra dipole trap that breaks
the discrete translational symmetry of the condensate.  The quasi
momentum $q$ is no longer a good quantum number. The dimension of the
BEC along the cavity axis is typically around $10\,\mu{}m$, which is
around $10$-$15$ wavelengths. As a result, the selection of the
optomechanical mode by the cavity mode is not perfect, and there can
be \emph{more} BEC excitation modes that couple to the cavity
field. Moreover, atom-atom collision becomes significant too
in shaping the condensate. Therefore the Bloch state expansion of the BEC
excitations is not suitable any more.  In the following, we
investigate what are the relevant optomechanical modes and how they
are composed of the discrete condensate excitation modes of a trapped
BEC.

\begin{figure}[t]
\centering
\includegraphics[width=0.8\columnwidth]{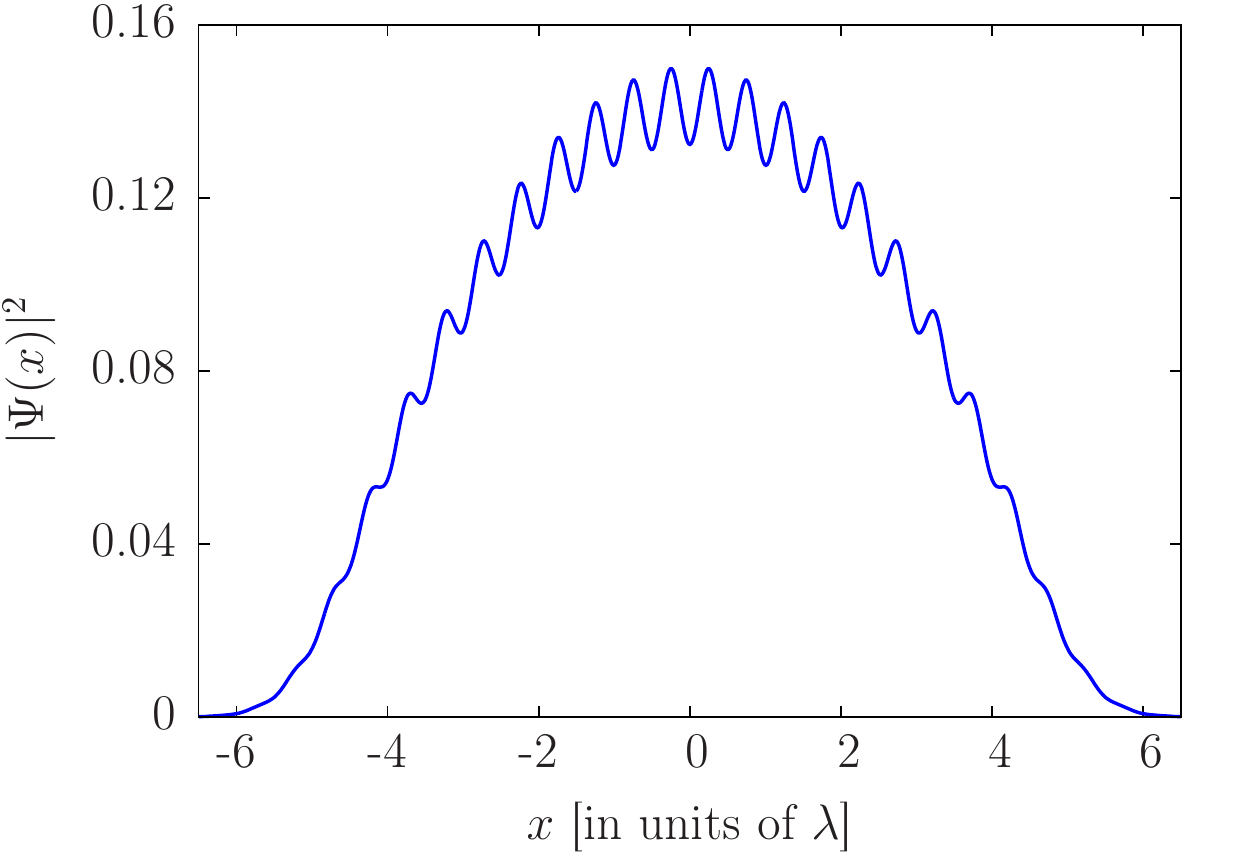}
\caption{The BEC wave function for $N=6\cdot 10^4$, $U_0 = 0.96$,
  $\eta=549.5$, $\delta_C = 0$, $gN=2$, $V_{tr} = 0.01$
  $\omega_R$. The corresponding Thomas-Fermi radius is $5.31$
  $\lambda$.  }
\label{fig:psi}
\end{figure}

We consider Eqs.~(\ref{eq:H_total}) and (\ref{eq:cavity}) in real
space using a mean-field model similar to the one described in
Sec.~\ref{subsec:mf_model}. We assume an external harmonic trap
potential for the atoms, $V_{\rm ext}(x) = V_{tr} x^2$.  Similarly to
the procedure of Sec.~\ref{subsec:mf_model}, we split the atomic field
to a condensate mean-field wave function $\psi(x)$ and fluctuations
$\delta\Psi(x,t)$, \be \Psi(x,t) = \sqrt{N}e^{-i\mu{}t}\left[\psi(x) +
  \delta{}\Psi(x,t)\right]\,.  \ee We solve the mean field equations
(Eqs.~(4a-b) of Ref.~\cite{Szirmai2010Quantum}) for a
condensate size of $6$ to $10$ $\lambda$ by discretizing the
coordinate on $800$ to $1400$ grid points. A typical wave function is
shown in Fig.~\ref{fig:psi}. For the selected parameters, the
condensate takes the shape of the parabolic trap, i.e., it is deeply
in the Thomas-Fermi regime. However, it is slightly modulated by the
cavity field potential. This modulated component of the wave function
corresponds to the $\cos\,2kx$ mode in the homogeneous case in
Eq.~(\ref{eq:mode_exp}). Here, the wave function can be approximated
by $\psi(x) = e(x) + \sqrt{2}\cos(2kx)f(x)$, where $e(x)$ and $f(x)$
are slowly varying functions on the scale of $\lambda$.

\begin{figure}[t]
\centering
\includegraphics[width=0.78\columnwidth]{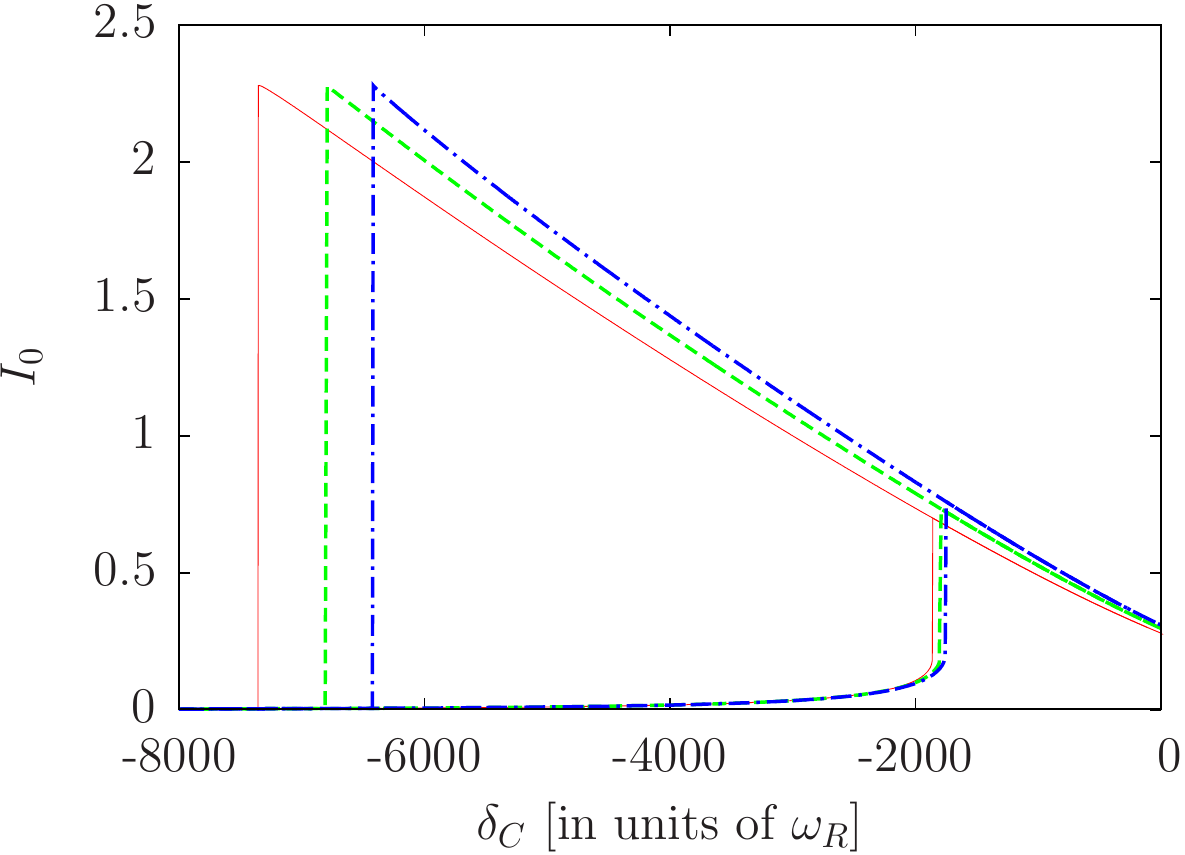}
\caption{The intracavity photon number $I_0$ vs the detuning parameter
  $\delta_C$ for a trapped BEC with Thomas-Fermi radii $3.11$
  $\lambda$ ($V_{tr} = 0.05$) (blue dashed dotted line) and $5.31$
  $\lambda$ ($V_{tr} = 0.01$) (green dashed line). The parameters are:
  $N=6\cdot 10^4$, $U_0 = 0.96$, $\eta=549.5$, $\kappa=363.9\omega_R$
  and $gN = 2$. The noninteracting limit ($gN = 0$) of the homogeneous
  condensate ($V_{tr}=0$) is plotted for reference (solid red line). }
\label{fig:phnum_tr}
\end{figure}

Obviously, the system exhibits similar bistable behavior as in the untrapped
case. In Fig.~\ref{fig:phnum_tr} we plot the cavity photon number as a
function of the cavity detuning $\delta_C$ for fixed $gN=2$ but for
two different trap frequencies $V_{tr}$. Note that for a trapped BEC,
the s-wave interaction energy depends on the size of the condensate.
In contrary to the homogeneous case, the parabolic trapping potential introduces a length scale and controls the condensate density and ultimately the s-wave scattering. When $V_{tr}\rightarrow0$ with a fixed $N$ one reaches the noninteracting limit of a homogeneous BEC (depicted with solid red lines in
Fig.~\ref{fig:phnum_tr}). We need to keep in mind that the atom-photon interaction is proportional to the atom number and not to the density.

\begin{figure}[t]
\centering
\includegraphics[width=0.9\columnwidth]{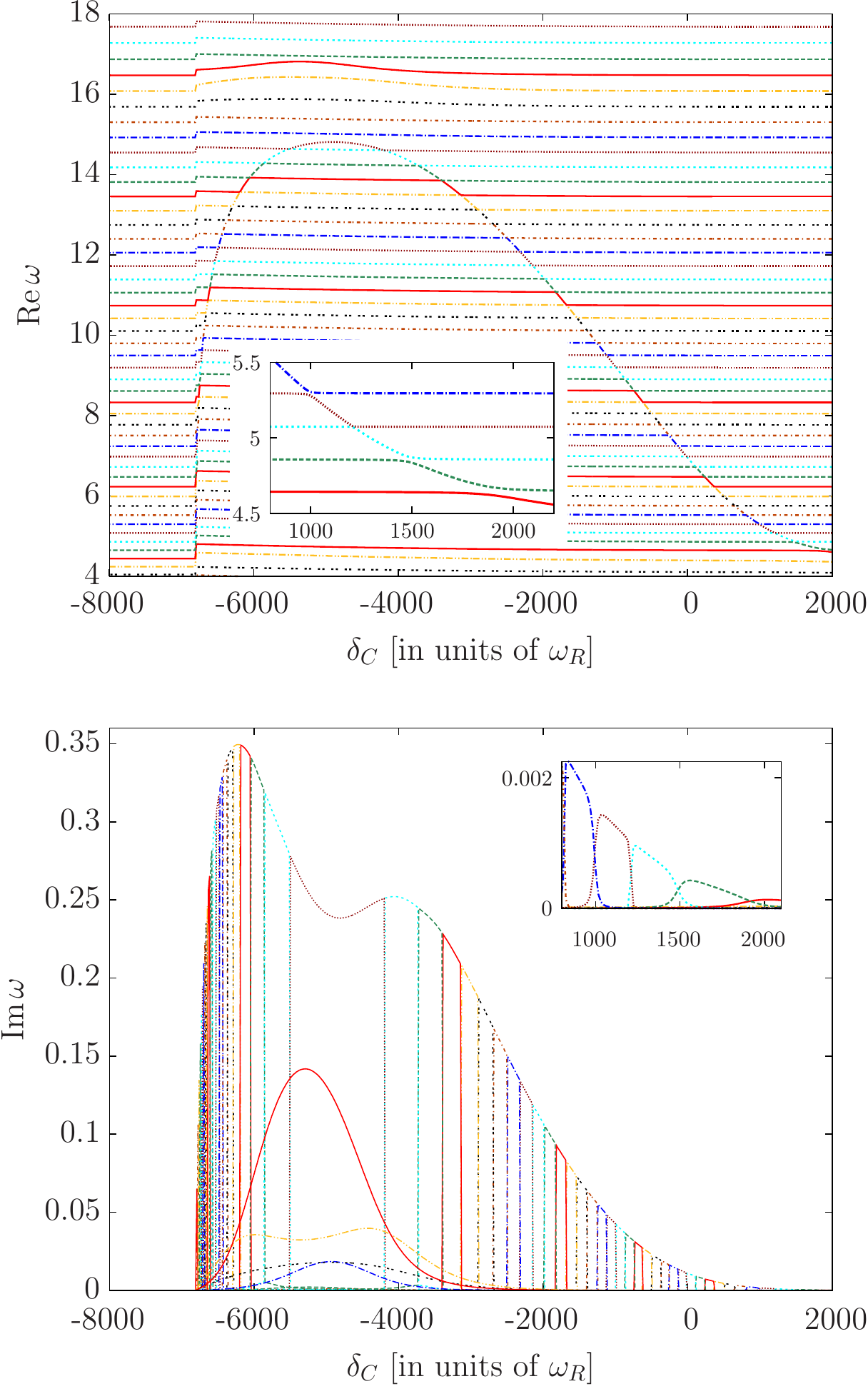}
\caption{Frequency (top) and decay rate (bottom) of the relevant
  optomechanical excitation modes as a function of the cavity detuning
  $\delta_C$ on the upper branch of the bistable regime (cf. with
  Fig.~\ref{fig:optomode}). The insets magnify the avoided crossings
  for the low lying excitation modes.  The plots comprise solely the
  modes with even parity symmetry.  The parameters are: $N=6\cdot
  10^4$, $U_0 = 0.96$, $\eta=549.5$, $\kappa=363.9$, $V_{tr} = 0.01$
  and $gN = 2$ $\omega_R$.}
\label{fig:optomodes_tr}
\end{figure}

In Fig.~\ref{fig:optomodes_tr}, we show the excitation spectrum above the mean-field along
the upper branch solution. Far off the
cavity resonance domain, {\it i.e.}, for $\delta_C<-8000\,\omega_R$ or
$\delta_C>2000\,\omega_R$, the spectrum exhibits the excitation levels
of a trapped condensate being approximately in the collision dominated
Thomas-Fermi regime
\cite{Csordas1999Collective,Menotti2002Collective}. One feature of the
plot is that these levels represent an inert background of almost
horizontal lines throughout the detuning range considered.  There is a
jump in these lines at $\delta_C=-6800\,\omega_R$ which corresponds to
the stability threshold of the upper branch. Below this detuning the
mean field solution must be on the lower branch of
Fig.~\ref{fig:phnum_tr} with less cavity intensity and thus less
perturbation of the external trapping potential $V_{\rm ext}$. The
main feature of this plot is that the optomechanical mode appears clearly and  
its spectral line resembles very much to that of the homogeneous system shown in
Fig.~\ref{fig:optomode}.  This relevant optomechanical mode grows out from the one which has about $4.47\,\omega_R$ frequency without cavity-BEC interaction, e.g., in the far
detuned regime given by the left and right extremes of the plot. Its
wave function overlaps the most with the cavity mode function, hence
it dominantly couples to the cavity field fluctuations. In the
resonant regime $-6800 < \delta_C/\omega_R < 2000$ the excitation
frequency of this mode crosses the discrete energy levels of the
trapped condensate. More precisely, the coupling of the other modes to
the cavity field, though negligibly small, is not exactly
zero. Therefore what one can see is a series of avoided crossings (at
these points the lines associated with modes swap color). Most of the
avoided crossings cannot be numerically resolved. At the sides of the plot, some well resolved avoided crossings show
that more than one condensate excitation couples considerably to the
cavity. Such an example is shown in the inset of the Figure at around
$\delta_C\approx 1500\,\omega_R$. 

The condensate size defines the overlap between the BEC excitation
modes and the cavity mode. In Fig.~\ref{fig:optomodes_tr} a trapped
BEC of Thomas-Fermi radius $5.31\lambda$, {\it i.e.}, about 10 times
the optical wavelength leads to a good enough single mode selection
and to a behavior similar to that of a homogeneous gas. If the size
was reduced, the modes neighboring the optomechanical mode would have
larger overlap with the cavity mode. However, their separation in
energy would also increase, reducing their effective coupling to the
optomechanical dynamics. Reversely, for larger condensate the detuning
of the adjacent modes is smaller but the geometrical single mode
selection becomes more precise. Finally, note that modes with odd
parity do not couple to the cavity field and are not plotted in
Fig.~\ref{fig:optomodes_tr} at all.

There is an efficient way to identify the condensate excitation modes
which couple to the cavity field. These ones have non-vanishing imaginary
part in the spectrum (bottom panel of Fig.~\ref{fig:optomodes_tr})
which is a measure of the coupling.  The inset, corresponding to the
detuning range selected in the upper panel, clearly demonstrates that
two modes are relevant in the narrow vicinity of crossings. Compared
to the imaginary part of the spectrum shown in Fig.~\ref{fig:optomode} for
homogeneous gas, the envelope indicating the decay rate of the
optomechanical mode has a different shape here. In particular, there
is a dip at $\delta_C=-4800\,\omega_R$ that can be attributed to the
contribution of higher excitation modes. In the homogeneous case, we
restricted the Hamiltonian to the lowest two bands, however in the
real-space solution all bands are present up to a cutting frequency
determined by the discretization. For a homogeneous BEC, the next
relevant contribution comes from the $n=2$ band in
Eq.~(\ref{eq:mode_exp}), since the cavity mode couples the $\cos(2kx)$
mode to the $\cos(4kx)$ one. This effect can be clearly recognized
here in Fig.~\ref{fig:optomodes_tr} in the form of a small modulation
of the excitation frequencies at $\delta_C\approx -5200\,\omega_R$ and
around the kinetic energy of the $\cos(4kx)$ mode ($16\,\omega_R$). The
mostly affected excitation mode gains a significant decay rate from
the cavity interaction (solid red peak at the same detuning).

\section{Conclusion}
\label{sec:conclusion}

A Bose-Einstein condensate, dispersively coupled to the field of a
laser-driven high-\emph{Q} cavity constitutes an alternative system to
study cavity optomechanics. Instead of a nanomechanical oscillator,
here, a single matter-wave excitation mode couples to the light
field, that is selected by the cavity mode function. With this system,
a new regime of cavity optomechanics can be studied experimentally. 

In this paper, we included s-wave atom-atom collision in the
optomechanical model and showed its effects on the optical
bistability of the system. For a homogeneous BEC, we derived a
radiation pressure Hamiltonian, from which we concluded that
collisions increased the frequency of the mechanical oscillator mode
and decreased the optomechanical coupling strength. As a result, the
bistable regime shrinks.

Due to collisions,  a spatially confined gas of degenerate ultracold atoms takes the parabolic profile
of the harmonic trap potential. In this case, more than one BEC
excitation modes can have finite overlap with the cavity mode function. There is no {\it 'a priori'} known
excitation mode that would represent a single mechanical oscillator. The calculated 
excitation spectrum reflects, however, that the single mode
picture applies in most of the cases, except for narrow ranges of weak avoided crossings. The contribution of modes
around the second harmonics of the cavity mode function was also
revealed from the real-space solution.

This work was supported by the Hungarian Academy of Sciences (Lend\"ulet Program, LP2011-016) and from the Hungarian National Office for Research and Technology (ERC\_HU\_09 OPTOMECH). G.~Szirmai acknowledges support from the Hungarian National Research Fund (OTKA T077629) and from the J\'anos Bolyai Scholarship.

\section*{Appendix}

In the Appendix, we give the explicit form of the matrices appearing
in Eqs.(\ref{eq:fluct_q}) and (\ref{eq:fluct_0}). Let us start with
the fluctuations of atomic operators with $q\neq0$. The equations of
motion are derived from the Heisenberg equations, $i\hbar\partial_t
b_q=[b_q,H]$ (and similarly for $b_q^\dagger$), with the Hamiltonian
\eqref{eq:H_quantized}, by including the mean-field expansion
Eq.~(\ref{eq:mf_expansion}), and linearizing in the operators. The
photon field do not couple to these fluctuations since it carries zero
quasimomentum. For the ${\bf L}_q$ matrix, appearing in the equation
of motion \eqref{eq:fluct_q}, with a straightforward algebra one
obtains:
\begin{subequations}
\be
\bmx
\omega_q^b & {\cal G} & {\cal B} & {\cal C} & 0 & 0 \\
-{\cal G} & - \omega_q^b & -{\cal C} & -{\cal B} & 0 & 0 \\
{\cal B} & {\cal C} & \omega_q^c & {\cal G}(1 + \frac{\gamma_0^2}2) & 4iq & 0 \\
-{\cal C} & -{\cal B} & -{\cal G}(1 + \frac{\gamma_0^2}2) 
& -\omega_q^c & 0 & -4iq \\
0 & 0 & -4iq & 0 & \omega_s^q & \frac{\cal G}{2}(1 + \beta_0^2) \\
0 & 0 & 0 & 4iq & -\frac{\cal G}{2}(1 + \beta_0^2) & -\omega_s^q
\emx,
\ee
with
\bea
\omega_q^b &=& \omega_R q^2 + 2{\cal G} - \mu \,, \\
\omega_q^c &=& \omega_R (4 + q^2) + 2{\cal G} - \mu \,, \\
{\cal C} &= & 2{\cal G}\beta_0\gamma_0 \,,\\
{\cal B} &= & u |\alpha|^2 + 2{\cal C} \,.
\eea
\end{subequations}
The equations of motion for the fluctuations of atomic operators with $q=0$ are also obtained directly from their Heisenberg equations and the mean-field substitution \eqref{eq:mf_expansion}, but now Eq.~\eqref{eq:cavity} is also needed since the photon field couples to these polariton like modes.
Finally, the matrix {\bf M} of Eq.~\eqref{eq:fluct_0} reads,
\begin{subequations}
\be
\bmx
{\cal A} & 0 & u\alpha\gamma_0 & u\alpha\gamma_0 & u\alpha\beta_0  & u\alpha\beta_0 \\
0 & -{\cal A^{*}} & -u\alpha^{*}\gamma_0 & -u\alpha^{*}\gamma_0 & -u\alpha^{*}\beta_0 & -u\alpha^{*}\beta_0 \\
u\alpha^{*}\gamma_0 & u\alpha\gamma_0 & \omega_b & {\cal G} &
{\cal B} & {\cal C} \\
-u\alpha^{*}\gamma_0 & -u\alpha\gamma_0 & -{\cal G} & -\omega_b &
-{\cal C} & -{\cal B} \\
u\alpha^{*}\beta_0 & u\alpha\beta_0 & {\cal B} & {\cal C} & \omega_c &
{\cal G}(1 + \frac{\gamma_0^2}2) \\
-u\alpha^{*}\beta_0 & -u\alpha\beta_0 & -{\cal C} & -{\cal B} & -{\cal
G} (1 + \frac{\gamma_0^2}2) & -\omega_c
\emx
\ee
with
\bea
{\cal A} &=& -\delta_C + 2\beta_0\gamma_0Nu - i\kappa\,,\\
\omega^b &=& 2{\cal G} - \mu\,,\\
\omega^c &=& 4\omega_R + {\cal G}(2 + \gamma_0^2) - \mu\,.
\eea
\end{subequations}

\end{document}